\documentclass[superscriptaddress,showpacs,amssymb,10pt,aps,prd,reprint,longbibliography,footinbib]{revtex4-1}
\usepackage{graphicx,epsfig,amssymb,times} 
\usepackage{amsmath,amsfonts}
\usepackage{bm}
\usepackage{epstopdf}
\usepackage[linktocpage,colorlinks]{hyperref}
\usepackage[caption=false]{subfig}
\usepackage[usenames]{color}     
\usepackage{float}
\usepackage{natbib}
\usepackage{ulem}
\usepackage{cancel}
\usepackage{verbatim}
\usepackage{enumitem}
\usepackage[utf8]{inputenc}
\definecolor{coolblack}{rgb}{0.0, 0.18, 0.39}
\definecolor{darkred}{rgb}{0.5,0,0}
\definecolor{darkgreen}{rgb}{0,0.5,0}
\definecolor{darkblue}{rgb}{0,0,0.5}
\definecolor{lapislazuli}{rgb}{0.15, 0.38, 0.61}
\definecolor{venetianred}{rgb}{0.78, 0.03, 0.08}
\definecolor{bleudefrance}{rgb}{0.19, 0.55, 0.91}
\definecolor{dogwoodrose}{rgb}{0.84, 0.09, 0.41}
\hypersetup{colorlinks=true, citecolor=darkblue, linkcolor=darkblue, 
urlcolor = darkblue}
\begin{document}


\title{Particle collisions near static spherically symmetric black holes}


\author{Eva Hackmann}\email{eva.hackmann@zarm.uni-bremen.de}
\affiliation{ZARM, University of Bremen, Am Fallturm, 28359 Bremen, Germany.}

\author{Hemwati Nandan}\email{hnandan@associate.iucaa.in}
\affiliation{Department of Physics, Gurukul Kangri Vishwavidyalaya, Haridwar-249 407, India.}
\affiliation{Center for Space Research, North-West University, Mafikeng 2745, South Africa.}

\author{Pankaj Sheoran}\email{hukmipankaj@gmail.com}
\affiliation{Instituto de F\'{\i}sica y Matem\'{a}ticas, Universidad Michoacana de San Nicol\'{a}s de Hidalgo,\\
Edificio C-3, 58040 Morelia, Michoac\'{a}n, M\'{e}xico.}


\date{\today}

\begin{abstract}
It has been shown by Ba\~{n}ados, Silk and West (BSW) that the center of mass energy ($E_{\rm cm}$) of test particles starting from rest at infinity and colliding near the horizon of a Schwarzschild black hole is always finite. In this short note, we extent the BSW scenario and study two particles with different energies colliding near the horizon of a static spherically symmetric black hole. Interestingly, we find that even for the static spherically symmetric (i.e., Schwarzschild like) black holes it is possible to obtain an arbitrarily high $E_{\rm cm}$ from the two test particles colliding near but outside of the horizon of a black hole, if one fine-tunes the parameters of geodesic motion.
\end{abstract}


\maketitle

\section{Introduction}
In 2009, Ba\~{n}dos, Silk and West (BSW) first showed \cite{Banados:2009pr} that an extremal axially symmetric rotating black hole (BH) can act as a particle accelerator: an extremely high amount of center of mass energy ($E_{\rm cm}$) can be produced from the collision of two test particles starting from rest at infinity. This is in contrast to static spherical symmetric BHs where $E_{\rm cm}$ is always finite. They pointed out that the extremal rotating BHs may therefore be used as an important probe of high energy scale physics. Since then collisions of geodesic (and charged) particles near the horizon of legion of BHs \cite{Zaslavskii:2010jd,Zaslavskii:2010pw,Wei:2010vca,Wei:2010gq,Kimura:2010qy,Harada:2011xz,Galajinsky:2013as,Ghosh:2014mea,Amir:2015pja,Ghosh:2015pra,Zakria:2015eua,Amir:2016nti,Gonzalez:2018lfs,Ogasawara:2018gni} has been analysed. The study of BSW effect is not limited to geodesic particles only but extended to spinning and accelerated particles as well \cite{Armaza:2015eha,Zaslavskii:2016dfh,Zaslavskii:2019bho}. However, all these different cases of collisions of  particles (i.e., geodesic, charged, spinning, and accelerated) in the vicinity of a BH are accompanied by some serious limitations. For the case of geodesic particles, extremely high $E_{\rm cm}$ is obtained if the collision occurs near the horizon of an extremal rotating BH and the energy $E$ and the angular momentum $L$ of one of the colliding particles meet a critical condition \cite{Banados:2009pr}. On the other hand, when the collision of spinning particles is considered far away from the event horizon of a BH, the main difficulty is the unavoidable occurrence of a superluminal region (i.e. the region where the four velocity of the spinning particle changes from timelike to spacelike) \cite{Zaslavskii:2016dfh}. Finally, one can obtain the arbitrarily high $E_{\rm cm}$ even for the static spherically symmetric case, but it requires that one of the colliding particles must be accelerated, which in turn requires the existence of some external source (i.e. which act as an engine to accelerate the particle) \cite{Zaslavskii:2019bho}. With
these limitations on the BSW effect, it is natural to investigate whether a BH in its simplest form (i.e. static spherically symmetric BH) can act as an accelerator for colliding geodesic particles. 

\begin{figure}[h!]
\hspace*{-0.0cm}
\includegraphics[scale=0.48]{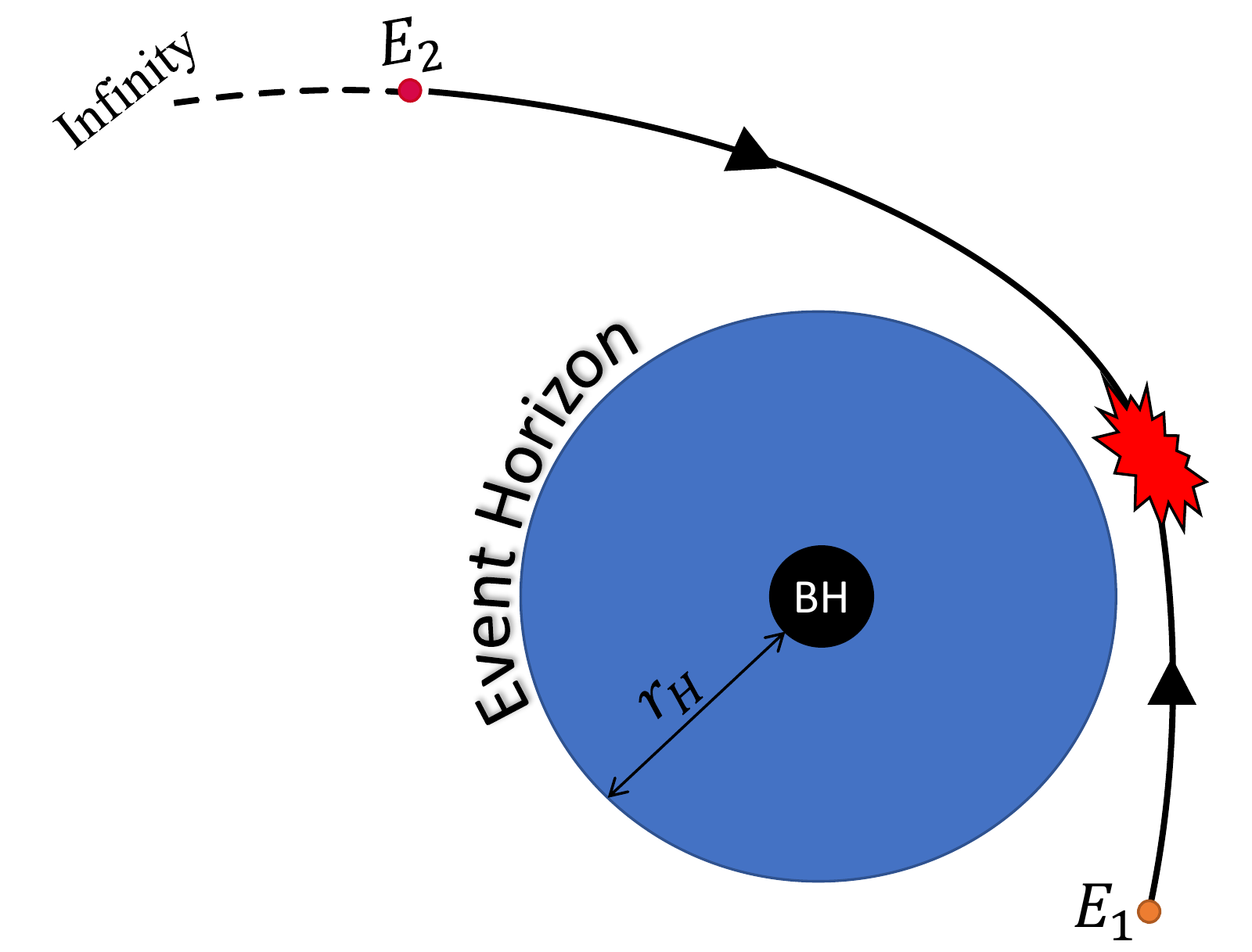}
\vspace{0cm} 
 \caption{Schematic view of two geodesic particles with different energies ($E_{1}\neq E_{2}$) colliding near the horizon of static and spherically symmetric BH.}
\label{fig1}
\end{figure}

In this paper, we study a collision of two particles near the horizon of a static spherically symmetric BH. We assume that one of the colliding particles originates near the horizon of the BH and posses very small radial motion with energy such that $E_{1} = \epsilon \ll 1$. On the contrary, the second particle is generic, for instance falling from rest at infinity with energy $E_{2}$ as shown in Fig. \ref{fig1}. This is a new situation for the colliding particles that has not been studied yet, and some novel features of particle acceleration process in the vicinity of a BH are reported.  

The paper is organized as follows. In Sec. \ref{sec:Collision_gp}, we briefly describe the metric of the static spherically BH, the equation of motion of the geodesic particles, and the expression for the center of mass energy $E_{\rm cm}$. In Sec. \ref{sec:Arbitrarily_high_Ecm}, we explain how a static spherically symmetric BH can act as a particle accelerator for two geodesic particles colliding in its vicinity. Finally, in Sec. \ref{sec:Conclusion}, we summarize our main results with a concluding remark. Throughout this paper, we choose the $(-,+,+,+)$ signature for the metric tensor, Greek indices run from $0$ to $3$, and both the fundamental constants $G$ and $c$ are set equal to unity.

\section{Collision of geodesic particles}
\label{sec:Collision_gp}
A static and spherically symmetric spacetime is given by,
\begin{align}
g & = -f(r) dt^2 + \frac{dr^2}{f(r)} + r^2(d\theta^2 + \sin^2\theta d\phi^2)\,.
\end{align}
We assume that we have a BH solution with an event horizon located at $r=r_H$. Dependent on the choice of the metric function $f(r)$ we may have additional horizons, for instance a cosmological or a Cauchy horizon. The two Killing vectors $u_t = -E$ and $u_\phi = L$ thus give rise to two constants of motion as below,
\begin{align}
-E & = g_{tt} \dot{t}\,,\\
L & = g_{\phi\phi} \dot{\phi}\,,
\end{align}
where a dot denotes a derivative with respect to the proper time of the particle. Due to spherical symmetry, without loss of generality, we entirely restrict to the plane $\theta = \pi/2$. From the normalization condition $g_{\mu\nu} \dot{x}^\mu \dot{x}^\nu = -1$, one can then obtain the following equation of geodesic motion,
\begin{align}
\dot{r}^2 & = E^2 - f(r) \left( 1+ \frac{L^2}{r^2} \right) =: R(r)\,.\label{eq:r}
\end{align}  

Let us now consider two particles of masses $m_1$ and $m_2$ moving on geodesics in a BH spacetime. Their energy in the center of mass frame is given by,
\begin{align}
E_{\rm cm}^2 & = -g_{\mu\nu} (p_1^\mu + p_2^\mu) (p_1^\nu + p_2^\nu)\\
& = m_1^2 + m_2^2 - 2g_{\mu\nu} p_1^\mu p_2^\nu\,,
\end{align}
where $p_i^\mu$ denotes the four momentum of the two particles. If we assume $m_1=m_2=m$, one can easily obtain, 
\begin{align}
\frac{E_{\rm cm}^2}{2m^2} & = 1 - \frac{L_1L_2}{r^2} + \frac{E_1E_2-\sqrt{R_1(r)R_2(r)}}{f(r)}\,,\label{Ecm}
\end{align}
where the indices refer to the two different particles. Regardless of the apparent singularity at the horizon, where $f(r_H)=0$, for generic particles the center of mass energy remains finite as the nominator also vanishes at $r=r_H$. The limit is given by,
\begin{align}
\lim_{r \to r_H} \frac{E_{\rm cm}^2}{2m^2} & = 1 - \frac{L_1L_2}{r_H^2} + \frac{E_1}{2E_2} \left( 1+ \frac{L_2^2}{r_H^2} \right)\nonumber\\
& \quad + \frac{E_2}{2E_1} \left( 1 + \frac{L_1^2}{r_H^2} \right)\,. \label{Ecmlimit}
\end{align}
In the original paper by BSW, it was assumed from the beginning that the two particles start from rest at infinity, implying $E_1=E_2=1$. In this case, we see that the center of mass energy is always finite. Later, it was pointed out by Zaslavskii that for charged particles in a Reissner-Nordström spacetime, or geodesic particles in a stationary (but non-static) and axially symmetric spacetime the parameters can be fine-tuned such that the expression Eq. \eqref{Ecmlimit} becomes arbitrarily large. For the Schwarzschild spacetime, Grib et al \cite{Grib2012} show that, in the limit $r \to r_H$ considered above, if the point of collision coincides with the radial turning point $\dot{r}=0$ of one of the particles than the energy may also grow without bound. We generalise here the argument from \cite{Grib2012}, allowing for general static spherically symmetric spacetimes, and removing the condition that the point of collision coincides with the radial turning point.
\vspace{0.2cm}

\section{Arbitrarily large center of mass energy}
\label{sec:Arbitrarily_high_Ecm}
We immediately notice from the limit Eq. \eqref{Ecmlimit} that the center of mass energy diverges if one of the particles has zero energy. However, as evident from eq. \eqref{eq:r}, this means that geodesic motion is impossible. Instead, we may consider a particle with a very low energy. Assume that particle 1 has $E_1 = \epsilon \ll 1$ and particle 2 is generic. Then from Eq. \eqref{Ecmlimit}, one can observe,
\begin{align}
\lim_{r \to r_H} \frac{E_{\rm cm}^2}{2m^2} & = \frac{E_2}{2\epsilon} \left( 1 + \frac{L_1^2}{r_H^2} \right) + \mathcal{O}(\epsilon^0)\,.
\end{align} 
This result was basically derived in \cite{Grib2012}, although we did not explicitly assume that $\dot{r}_1=0$ at the horizon. However, as mentioned above, geodesic motion with $E_1 \ll 1$ is only possible if $\dot{r}_1=0$ very close to the horizon. Note that if both particles have $E \ll1$, the center of mass energy becomes again finite.

More generally, we can also obtain arbitrarily large center of mass energy for collisions close to the horizon, but outside of it. Let $r_c$ be the point of collision with $r_c-r_H \ll 1$. We rewrite the metric function $f$ as $f(r) = (r-r_H) \tilde{f}(r)$, in particular $\tilde{f}(r)=1/r$ for Schwarzschild spacetime. Let us choose for, say particle 1, $E_1= C_1\sqrt{r_c-r_H}$. Here $C_1$ is a positive constant, taken such that geodesic motion is possible in the region $(r_H,r_c)$. We then find the following expression from Eq. \eqref{Ecm} at the point of collision $r_c$,

\begin{widetext}
\begin{align}
\frac{E_{\rm cm}^2}{2m^2} & = 1 -  \frac{L_1L_2}{r_c^2} + \frac{E_2C_1\sqrt{r_c-r_H}}{(r_c-r_H)\tilde{f}(r_c)} - \frac{1}{(r_c-r_H)\tilde{f}(r_c)} \times\nonumber\\
& \quad \times \left[ (r_c-r_H) \left(C_1^2 -\tilde{f}(r_c) \left(1+\frac{L_1^2}{r_c^2}\right) \right) \left(E_2^2 - (r_c-r_H) \tilde{f}(r_c) \left(1+\frac{L_2^2}{r_c^2}\right) \right)   \right]^{\frac{1}{2}}\\
& =  1 -  \frac{L_1L_2}{r_c^2} + \frac{E_2C_1}{\sqrt{r_c-r_H}\tilde{f}(r_c)} - \frac{E_2\sqrt{C_1^2-\tilde{f}(r_c)\left(1+\frac{L_1^2}{r_c^2}\right)}}{\sqrt{r_c-r_H}\tilde{f}(r_c)} + \mathcal{O}(\sqrt{r_c-r_H})\,.
\end{align}
\end{widetext}

Therefore, for small but nonvanishing $r_c-r_H$, we can obtain arbitrarily large center of mass energy provided we can choose the constant $C_1$ such that
\begin{align}
C_1 - \sqrt{C_1^2-\tilde{f}(r_c)\left(1+\frac{L_1^2}{r_c^2}\right)} >0\,.
\end{align} 
For instance, we could choose 
\begin{align}
C_1 = \sqrt{ 1 + \frac{\tilde{f}(r_c)(r_c^2+L_1^2)}{r_c^2}}\,.
\end{align}
Then, at the point of collision we find from Eq. \eqref{eq:r} that $\dot{r}^2|_{r=r_c}=(r_c-r_H)>0$ as well.

A caveat of both calculations explained above is that the particle 1 is assumed to have a very small radial motion velocity although it is very close to the horizon. This is a very special situation, which will usually not be encountered, and probably will only arise if the particle originated from a foregoing collision. 

\section{Summary and conclusion}
\label{sec:Conclusion}
In this short paper, we delved into the BSW effect and studied the collision of two geodesic particles near the horizon of a static spherically symmetric BH, generalising a scenario shortly discussed in \cite{Grib2012}. It is assumed that one of the colliding particles (say particle 1) originates near the horizon of the BH and posses a very small radial velocity and energy $E_{1}\ll 1$ (despite the fact that it is very close to $r_{H}$). On the other hand, the second colliding particle with energy $E_{2}$ is considered to be generic, for instance the usual particle coming from rest at infinity. It is shown that when the collision of the above mentioned geodesic particles takes place near the event horizon of a static spherically symmetric BH, it is still possible to have an arbitrarily high $E_{\rm cm}$. This is in contrast to the case when the colliding particles needs to have a spin as pointed out in \cite{Armaza:2015eha,Zaslavskii:2016dfh}, a charge as considered in \cite{Zaslavskii:2010aw}, or to the case when one needs an external engine (i.e. source) to accelerate one of the colliding particles in order to achieve arbitrarily high $E_{\rm cm}$ for static spherically symmetric BHs as pointed out in \cite{Zaslavskii:2019bho}. 

It is noteworthy to mention that even a static spherically symmetric BH can also act as a particle accelerator for geodesic particles and one do not need special types of particles (i.e. spinning or accelerated) in order to obtain arbitrarily high $E_{\rm cm}$. However, the scenario we discussed here is very special and is only possible if the particle 1 is originated near to horizon of BH by some process, maybe through a forgoing collision of particles.

In response to an earlier version of this paper, Zaslavskii \cite{Zaslavskii2020} analysed the physical relevance of the scenario under discussion here. He found that the for the Schwarzschild spacetime it is unphysical, requiring infinite forces to place the critical particle near the horizon. However, for the extremal Reissner-Nordstr\"om spacetime the critical (neutral) particle could be produced by a foregoing collision of charged particles. For other static and spherically symmetric spacetimes the question of physical relevance remains open.

\begin{acknowledgments}
 HN would like to thank Science and Engineering Research Board (SERB), New Delhi, India for financial support through grant no. EMR/2017/000339. HN is also thankful to IUCAA, Pune, India (where a part of the work was completed) for support in form of academic visits under its Associateship programme. PS would like to thank \textit{Programa de Desarrollo Profesional Docente} (PRODEP) of the \textit{Secretar\'{\i}a de Educac\'{\i}on P\'{u}blica} (SEP) of the Mexican government, for providing the financial support. EH is thankful for support from the research training group RTG 1620 "Models of Gravity" and the Cluster of Excellence EXC-2123 Quantum Frontiers - 390837967, both funded by the Deutsche Forschungsgemeinschaft (DFG, German Research Foundation).
\end{acknowledgments}

\bibliography{my.bib}
\bibliographystyle{apsrev}

\end{document}